\documentclass[preprint2]{aastex}
\usepackage{graphicx}
\shorttitle{Quasar Structure and Cosmological Feedback}
\shortauthors{Martin Elvis}

\begin{document}
%
%
\title{Quasar Structure and Cosmological Feedback}
%
%
\author{Martin Elvis}
%
\affil{Harvard-Smithsonian Center for Astrophysics
60 Garden St., Cambridge MA 02138 USA}
%
%
%
\begin{abstract}
Feedback from quasars and AGNs is being invoked frequently in several
cosmological settings.  Currently, order of magnitude, or more,
uncertainties in the structure of both the wind and the 'obscuring
torus' make predictions highly uncertain.  To make testable models of
this 'cosmological feedback' it is essential to understand the
detailed structure of AGNs sufficiently well to predict their
properties for the whole quasar population, at all redshifts.
Progress in both areas is rapid, and I describe the near-term
prospects for reducing these uncertainties for 'slow'
(non-relativistic) AGN winds and the obscuring torus.

\end{abstract}

\section{Introduction}

Feedback is the key to an interesting Universe. In particular,
feedback from quasars and their less luminous cousins, Active Galactic
Nuclei (AGNs), has been newly recognized as a potentially crucial
input to multiple areas of galaxy formation (\S\ref{feedback}).
However, so far, arguments using AGN feedback have been forced to make
simple assumptions: that all SMBHs accrete at the Eddington limit
while active and that 10\% the accreted mass is successfully ejected.

To become a testable science, ``cosmological feedback'' from AGNs must
use the details of the structure of quasar nuclei, both on a small
scale, where the winds most likely arise, and on a larger scale, at
the 'obscuring torus' of Unification Models \cite{Urry95}. 
%
%

\section{Cosmological Feedback}
\label{feedback}

There are six areas where AGNs are being called upon to provide
cosmological feedback:

\noindent{\bf 1. Co-evolution of SMBHs and their Host Galaxies.}  Some
form of feedback is required by the {\small $M_{BH}-\sigma_{BULGE}$}
relation \cite{Ferrarese00,Gebhardt00} so that the central black hole
does not grow at a rate independent of the surrounding dark matter
halo \cite{Silk98}.

\noindent{\bf 2.  Prevention of Star Formation in Mergers.}  The deep
HST GEMS survey \cite{Bell06} does not find the predicted blue branch
of young stars in the ($g-i$ vs. $M_V$) plane among the most massive
($M>2\times 10^{10}M_{\odot}$) galaxies, implying that star formation
is prevented during the mergers that form these galaxies ('dry
mergers'). Can AGN remove the cold ISM from these galaxies?

\noindent{\bf 3. Limiting the Upper Mass of Galaxies} $\Lambda$CDM
models produce too many high mass galaxies, contrary to
observations. Reduced cooling and feedback from supernovae are
insufficient to prevent galaxy growth \cite{Thoul95}. Heating by
AGN radio sources is a promising alternative \cite{Croton05}.

\noindent{\bf 3.  Inhibition of cooling flows} is demonstrated by {\em
XMM-Newton} spectra and {\em Chandra} imaging of the hot intracluster
medium of rich clusters of galaxies which show that their dense cores
are not cooling, and so not inflowing onto the central galaxy,
contrary to hydrostatic equilibrium model predictions
\cite{Kaastra01}. Instead something is providing an extra heat
source; very likely relativistic jets (\S\ 3.2).

\noindent{\bf 4.  Enrichment of the intergalactic medium.} Both the cool
Lyman-$\alpha$ forest \cite{Pettini04}, and the hotter `Warm-Hot
Intergalactic Medium' that produces the 'X-ray forest' \cite{Fang02}
seen in {\em Chandra} grating spectra \cite{Nicastro05} are far from
having a primordial composition, but are instead enriched with heavy
elements. Supernova driven 'superwinds' from starburst galaxies and
AGN winds can both escape their galaxies: which dominates in IGM
enrichment? 

\noindent{\bf 5. Dust at high redshift} is seen in z$\sim$6 quasars
\cite{Omont01}. Dust is important to catalyze efficient star formation
by shielding gas from UV heating and by enhancing cooling
\cite{Hirashita02}. But dust at z$\sim$6 is hard to make
\cite{Edmunds98}, and cannot be created in AGB-star winds (the
process that dominates in the Milky Way), as these stars take
$\sim$1~Gyr to evolve. Supernovae may create dust, but the rate is
unknown, so the origin of high z dust is open. Cool clumps in AGN
winds may be an effective alternative site.

\section{Pathways for AGN Feedback}

Quasars and AGN have three pathways by which they can provide
feedback:

\noindent {\bf 1. Radiation:} The defining characteristic of an AGN is
the huge radiative output, which can be comparable to that of an $L^*$
galaxy. This radiation carries energy and momentum that can affect the
quasar's environment. Radiation can enrich the IGM indirectly by {\em
heating, ionizing and accelerating the ISM from the quasar host
galaxy}, which inhibits star formation in the host. But
radiation is easily absorbed by dusty nuclear material.

Radiation pressure may be particularly important in a proposed
evolutionary phase when the quasar may blow away a layer of shrouding
material surrounding the SMBH at early epochs \cite{Sanders88}.

\noindent {\bf 2. Relativistic Jets:} Tightly collimated jets with
relativistic bulk velocities ($\Gamma\sim$10) commonly emanate from
the central galaxy in rich clusters of galaxies.  {\em Chandra} X-ray
images of the hot intracluster medium show holes into which the radio
structures fit like jigsaw pieces \cite{McNamara05}.  There is
clearly a close interaction between the relativistic plasma and the
X-ray hot plasma in these clusters. Even a tightly collimated jet will
spread heat throughout the intracluster medium and so {\em prevent a
cooling flow} \cite{Ruszkowski04}, so {\em setting an upper bound to
galaxy, and black hole, masses}. Only the most powerful jets, though,
escape their clusters to {\em enrich the IGM}.

However, relativistic jets are not common among AGNs: only about 10\%
of AGNs are radio loud, either because radio jet formation is a
transient phase of black hole activity, or because only a few black
holes are ever able to form a jet. In either case the total amount of
energy and momentum available from radio jets is reduced by this
factor, so that they may have difficulty solving other feedback
problems, especially in less massive systems.

It is possible that the majority of SMBHs, which - at any one time -
are quiescent rather than active, have 'dark' (i.e. non-radiative)
jets that carry substantial mass, energy and momentum - explaining why
they do not radiate anywhere near to the Bondi rate \cite{Soria06}.

\noindent {\bf 3. Slow (non-relativistic) Winds:} Moderate velocity
($\sim$1000 - 2000 km~s$^{-1}$) outflows are seen in absorption
$\sim$50\% of AGNs and quasars, and so form weakly collimated, wide
angle winds.  They are seen through the blueshifted narrow absorption
lines (NALs) they imprint on the UV continuum, and the `Warm Absorber'
(WA) features on the X-ray continuum. Less common ($\sim$15\%) are the
$\sim$10 times faster, but still non-relativistic, outflows seen in
the `Broad Absorption Line', BAL, quasars \cite{Crenshaw03}. Being
nearly universal in AGNs, these slow winds could create {\em
co-evolution}. BALs, at least, escape their hosts and {\em enrich the
IGM}.

Moreover, conditions in quasar winds at large radii, assuming that the
cooler ($\sim$10$^4$K), denser (10$^{10}$-10$^{11}$cm$^{-3}$), broad
emission line (BEL) gas is part of the wind, will match those in
AGB-star winds, and so dust should form copiously in AGN winds
\cite{Elvis02}, especially as high $z$ quasars have super-solar
abundances \cite{Hamann02}.

\subsection{Impulsive Events during Mergers}

An unexplored variant on the usually assumed continuous application of
these mechanisms over an AGN lifetime is an impulsive event
associated with a merger. Such an event could include all three
mechanisms. One object - the nearest radio-loud AGN: Cen~A/NGC~5128 -
gives us reason to consider this option. 

\begin{figure}[t!]
\resizebox{\hsize}{!}{\includegraphics[clip=true]{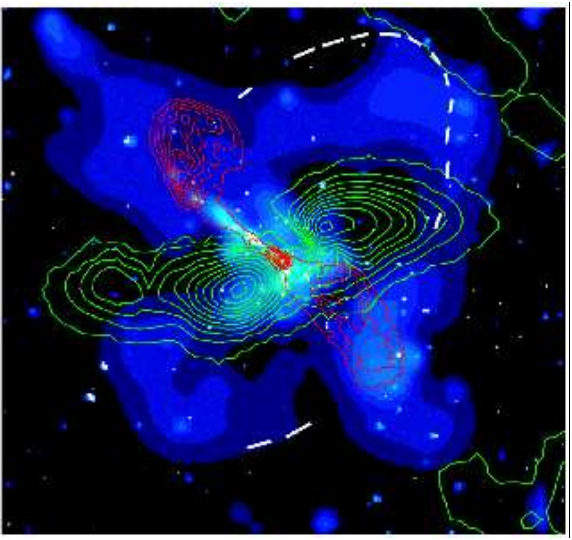}}
\caption{\footnotesize {\em Chandra} image of Cen~A/NGC~5128 (blue)
  overlaid with HI (green) and radio continuum (red) contours
  \cite{Karovska02}. An annulus of hot gas emission is evident
  perpendicular to the radio jet. This may be the result of an
  impulsive event $\sim$10$^7$yr ago, at the time of the galaxies'
  merger.}
\label{cena}
\end{figure}

Cen~A (fig.\ref{cena}) shows an elliptical 8~kpc radius annulus of hot
(kT$\sim$10$^6$~K) gas aligned with the radio jet axis
\cite{Karovska02}. This alignment seems to require a driving input
from the nuclear region. The thermal energy in this 'smoke ring' is
substantial, $\sim$10$^{55}$~erg, with a gas mass of
$\sim$10$^6$~M$_{\odot}$. Projecting backwards at the thermal
velocity, the ring would have been ejected $\sim$10$^7$~yr ago, about
the time of the evident merger in NGC~5128. Unfortunately such a
feature is visible, for now, only in Cen~A, because Cen~A lies just
3~Mpc away.

\section{Structure Influences Feedback}

Here I will first concentrate on slow, non-relativistic, winds, since
radiation and jet inputs have been long known, while the prevalence
and strength of slow AGN winds is only now becoming clear as a
potential source of cosmological feedback.  Also, progress here has
been rapid.  Then I will discuss how our ideas of the 'obscuring
torus' are developing away from the canonical 'donut', and how these
changes affect feedback.

\subsection{Slow Wind Structure}
\label{winds}

The mass loss rate in AGN winds, \.{M}$_W$, is uncertain by six
orders of magnitude.  This is because \.{M}$_W$ depends on the assumed
distance of the wind from the ionizing continuum source, $R$. For a
conical wind, with radial velocity $v_r$, and column density $N_H$
\cite{Krongold06}:
\begin{eqnarray}
\dot{M}_W = 0.8 \pi m_p N_H v_r R f(\delta,\phi)
\end{eqnarray}
[$f(\delta,\phi)$ is a factor that depends on the orientation of the
disk and the wind to our line of sight and, for reasonable angles, is
of order unity.]

The distance $R$ is uncertain by more than a factor 10$^6$ ($\sim 10$
kpc to $\sim 0.001$ pc). The proposed sites are: (a) the Narrow
Emission Line Region (NELR) \cite{Kinkhabwala02} (b) the inner edge
of the `obscuring torus' (\S\ref{torus}) \cite{Krolik95}, and (c) the
accretion disk itself \cite{Murrayetal95, Elvis00}. Large radii pose
a serious paradox for AGN winds: they require \.{M}$_{W} \sim$ 10 -
1000 \.{M}$_{acc}$, implying short-lived winds \cite{Netzer03}: a
result at odds with the high frequency of outflows in AGN.

Discriminating between these widely different scales requires breaking
the intrinsic degeneracy of the gas density, $n_e$, and distance from
the ionizing continuum source, $R$, in the equation that relates the
two observables: the luminosity of ionizing photons $Q_x$, and the
average ionization parameter of the gas
\begin{eqnarray}
U_X = Q_x / (4 \pi c R^2 n_e)
\end{eqnarray}

The answer is provided by time dependent photoionization \cite{Nicastro99}:
lower density gas recombines more slowly, so the lag between a
continuum change and the response of the WA $U_X$ can determine
$n_e$. Given the definition of $U_X$, $R$ follows, breaking the
degeneracy.  
%

\begin{figure}[t!]
\resizebox{\hsize}{!}{\includegraphics[clip=true]{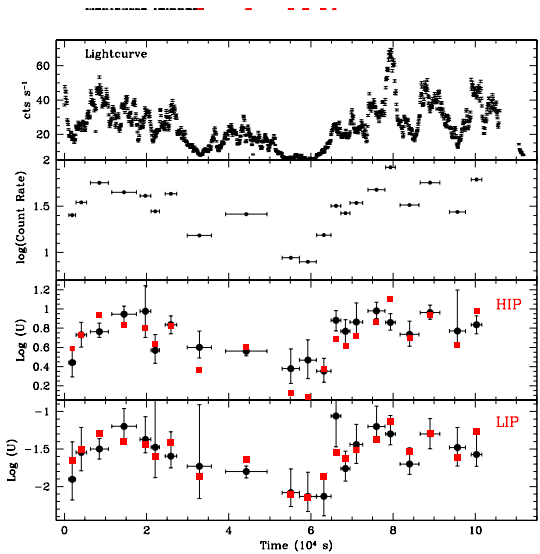}}
\caption{\footnotesize NGC~4051 {\em XMM} flux and $U_X$ light curves
for the two dominant WA components: the 'HIP' (High Ionization
Component) and 'LIP' (Low Ionization Component) \cite{Krongold06}.
The red squares show the predicted ionization parameter, $U_X$, for
ionization equilibrium.}
\label{lightcurve}
\end{figure}

\begin{figure}[t!]
\resizebox{\hsize}{!}{\includegraphics[clip=true]{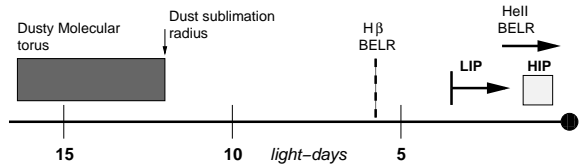}}
\caption{\footnotesize A map of the inner regions of the AGN NGC~4051.
The map is only 1-dimensional but is to scale, and shows the location
of the WA [high- ('HIP') and low-ionization ('LIP') components] in
NGC~4051 compared to the location of the high- (HeII) and
low-ionization (H$\beta$) broad emission line regions, and to the
innermost location of a dusty torus \cite{Krongold06}.}
\label{radii}
\end{figure}

We have recently applied this method to the WA in the narrow line
Seyfert 1 galaxy (NLSy1) NGC~4051, determining all the main physical
and geometrical properties of this WA \cite{Krongold06}.  The key to
success is the broad nature of the 0.6-0.9~keV Fe-M shell UTA and the
0.9-2~keV Fe-L shell and OVIII line complexes, which put strong
constraints on the $U_X$ of the two main WA components with XMM-EPIC
data. The high resolution/low signal-to-noise RGS spectrum provides
confirmation and sets the starting parameters, and both high- and
low-ionization (HIP, LIP) components follow the rapid variations of
the ionizing continuum (fig.\ref{lightcurve}).  Hence the WA gas must
be dense, and located at small radii, $R\sim few$1000~R$_S$,
i.e. accretion disk sizes. Because the BEL region (BELR) sizes in
NGC~4051 are also well known from reverberation mapping
\cite{Peterson00} we can draw a first map of the nucleus on a
well-determined physical scale (fig.\ref{radii}).

This result rules out a wind origin in a dusty obscuring torus, or any
larger region. Moreover, the derived wind radius is inside the
H$\beta$ emission line region, and is consistent with the high
ionization HeII emission line region size, long suspected to have an
outflowing component \cite{Gaskell82}. Several features of the
NGC4051 wind: the disk origin, high density, narrow thickness ($\Delta
R/R(HIP) \sim 0.1-0.2$), and pressure balance between HIP and LIP
\cite{Krongold06} are also features of my 'funnel-wind' model for
quasar structure \cite{Elvis00}, which suggests that something along
these lines will turn out to be the correct picture. If so, then we
have a tightly constrained geometry and kinematics which will admit of
few explanations.  We must then be close to a physical understanding
of AGN winds.

The implied mass outflow rate from the NGC~4051 wind is just $2-5$ \%
of \.{M}$_{acc}$, solving the \.{M}$_{W} >> $\.{M}$_{acc}$ paradox.
Yet, if this mass outflow rate is representative of all quasars,
powerful quasars still deploy large amounts of material and energy
into their environment.

However, NGC~4051 is a pathological AGN: with a small black hole mass
[(2$\times$10$^6$~M$_{\odot}$, \cite{Peterson00}], and the unusual
characteristics of NLSy1s (esp. rapidly variable X-rays with a steep
spectrum, and narrow broad emission lines, 'BELs' -and so a distant
BEL region, in $R_S$).  To extrapolate from NGC~4051 to all quasars at
all redshifts is risky. We need examples spanning the range of SMBH
masses and AGN luminosities.

The AGN winds seen as NALs are moving primarily transverse to our line
of sight \cite{Arav00, Mathur95}. Moreover if these outflows are
launched at $\sim$1000~R$_S$, as we find for NGC~4051
(\cite{Krongold06}, then the observed line-of-sight velocities are
well below escape velocity, so subsequent acceleration to BAL-like
velocities is required if the matter is not simply to fall back.  The
typical kinetic power in the slow wind is then increased by a factor
of order 100. We need to understand the driving physics of the wind to
know this factor.

\subsection{Obscuring Torus Structure}
\label{torus}

The longstanding picture of a dusty obscuring torus has served well,
but is now being re-assessed as new observations come in. I emphasize
that the basic insight of Unification Models is unchanged: angle
dependent obscuration clearly produces the two types of AGN
\cite{Lawrence82, deZotti85} in most cases [but see
\cite{Nicastro03}]. This was demonstrated convincingly by the finding
of 'hidden' BELs characteristic of 'type-1 AGN', in the polarized
spectrum in many otherwise narrow-lined 'type-2 AGN'
\cite{Antonucci85}.  The issue is what this torus consists of.

The obscuring torus also explains the observed 4:1 ratio of
type-2:type-1 AGNs, if the obscuring 'torus' covers 80\% of the AGN
sky. Feedback to larger scales will thus reduced by a factor 5 from
estimates based simply on SMBH mass density. The third feature
explained by an obscuring torus is the `ionization cone' structure
found on kiloparsec scales in several AGNs \cite{Tadhunter89}. The
torus can collimate the radiation to about the correct angle. Some
'ionization' cones though seem to be hollow expanding shells -
matter-bounded, not radiation-bounded \cite{Crenshaw00} - showing
large scale outflows at work.

The canonical picture of the obscuring torus \cite{Krolik88} is a that
of a 'donut' \cite{Urry95}: a dusty molecular ring that is Compton
thick (N$_H >$10$^{24}$cm$^{-2}$), and of large scale-height
(h/r$\sim$0.7), with an inner radius set by the dust sublimation
radius, $R_{sub}$, to be on $\sim$parsec scales \cite{Barvainis87}:
\begin{eqnarray}
R_{sub} = 1.3 L_{46}(UV)^{0.5}T(1500~K)^{-2.8} pc,
\end{eqnarray}
where $L_{46}(UV)$ is the ultraviolet continuum luminosity of the AGN
in units of 10$^{46}$erg~s$^{-1}$.

The longstanding problem with this picture is how to support a large
scale height in a cold structure. Thermal support is clearly out,
while invoking a mist of orbiting clouds will lead to flattening
through cloud-cloud collisions on an uncertain, but probably short,
timescale. A dynamic picture with continuous accretion onto the torus
from the host galaxy has more success \cite{Vollmer04}, but seems to
require \.{M}$_{torus} >$ \.{M}$_{Edd}$, which implies mass loss.

A more subtle difficulty with the 'donut' comes from the AGN 'photon
deficit' problem. It is a longstanding puzzle that the BELs from AGNs
emit more power than is present in the ionizing continuum
\cite{Netzer85} and, a related puzzle, require more ionizing photons
than are present by factors of 4 -- 25 \cite{Binette93}.  Variability
and dust extinction have been suggested to explain this deficit but do
not work well.  It seems that a major piece of the EUV continuum seen
by the BELR is missing from the observed spectrum.  How can the BEL
gas see a much stronger continuum than we do? If the BELR lies above
the disk [where the BELR would have to be part of the wind
\cite{Elvis00}], then the BELR sees the full UV radiation field from
the disk, while a typical observer ($i = 60^{\circ}$) sees the
continuum reduced by geometric and limb darkening factors
\cite{Netzer85}.  This may work, but a disk-aligned 'donut' would
prevent us seeing the BELR sufficiently edge-on to provide a large
enough effect.

The 'donut' picture is further complicated when AGN winds are
considered. The rapid acceleration expected from UV line driving
argues that BALs will be essentially equatorial \cite{Murrayetal95}.
Certainly, BALs must be at least partially radiatively accelerated as,
e.g., at least $\sim25$\% of the UV radiation emitted by the BAL
quasar PG~1254+047 is absorbed by a gas with
$N_H\sim$10$^{23}$~cm$^{-2}$ \cite{Hamann98}. So, if $L(UV)$ is even
10\% of the Eddington luminosity, the momentum in the wind is of the
order of that absorbed from the UV \cite{Risaliti06}.  BAL winds
cannot then lie in objects with co-aligned obscuring tori and
accretion disks, contrary to the usual assumption.

\smallskip

There are only a few ways out of these 'donut' related problems:
\begin{enumerate}
\item the wind may be polar or bi-conical and so rise above the torus;
\item the accretion disk and the torus may not be aligned;
\item the torus may be the host ISM;
\item the torus may be the wind.
\end{enumerate}

There is evidence that all four wind escape mechanisms occur:

\noindent {\bf 1. Bi-conical winds:} are indicated by transverse
motions and sub-escape velocities (see \S \ref{winds});

\noindent {\bf 2. Disk-torus misalignment:} If a radio jet axis shows
us the accretion disk orientation, and optical continuum polarization
position angle (PA) shows us the torus orientation, then early
evidence for their alignment \cite{Antonucci83} seems to be
supplanted by later, larger, samples \cite{Thompson88}.
Mis-alignment offers a solution to the photon deficit problem.

\noindent {\bf 3. Host galaxy obscuration:} Edge-on host galaxies have
a deficit of type~1 AGNs \cite{Keel80, Kirhakos90} indicating
obscuration related to the host \cite{Lawrence82}. Moreover the
optical polarization PA is aligned with the major axis of the host
galaxy disk \cite{Thompson88}, and obscuring kiloparsec scale dust has
been directly imaged in type~2 AGNs \cite{Malkan98}.  Even for the
archetype 'hidden type~1' AGN - NGC~1068 - CO imaging shows that it is
a warped disk on a $\sim$100~pc scale that blocks our view of the
nucleus \cite{Schinnerer00}. Variable X-ray obscuration is seen on a
timescale of a few years, suggesting similarly distant obscurers
\cite{Risaliti02}.

\noindent {\bf 4. Wind obscuration:} Large variations of the X-ray
obscuring column density within one day have now been seen in three
heavily obscured AGNs (N$_H \sim$10$^{22}$-10$^{23}$cm$^{-2}$)
\cite{Elvis04, Risaliti05, Puccetti06}. Such rapid changes can only
be accomplished, for material moving at Keplerian velocities, if the
matter lies close in, at about the BELR radius. This must be well
within the proposed torus, as the torus must hide the BELs, and would
be too hot for dust to survive at BELR radii.

This small-scale obscurer could be the wind, if the wind can be blown
off the accretion disk from radii where the disk temperature has
never risen above $\sim$1500~K, so that the material retains the
dust-to-gas ratio of the host galaxy ISM; a mix of this outer material
with more central hotter gas could explain the low dust-to-gas ratios
typically encountered in AGNs \cite{Maccacaro82, Maiolino01}.
Hydromagnetic models can reproduce the observed distribution of N$_H$
\cite{Kartje95, Kartje99}.

\smallskip
Clearly obscuration does take place on both a host galaxy ISM scale
and a quasi-BELR, slow wind, scale in many AGNs. A dusty disk wind is
toroidal and allows a large scale height obscuring region. As the
structure is a steady state flow, not a static structure, the torus
support problem disappears.

For cosmological feedback, the wind-as-torus picture still blocks 80\%
of the radiation from affecting the host galaxy, but the mass, energy
and momentum of the wind all escape cleanly. When the main obscuring
matter is located in the host galaxy, then the torus is the very
target ISM material that the AGN is supposed to heat, ionize and
remove, so all the radiative energy is also available. Both a
bi-conical wind and a misaligned disk and torus allow the wind to
escape, often to the IGM.

This is still an emerging picture, but shows promise. Without knowing
the details though, we will not be able to discriminate which is the
important element for feedback in particular circumstances, nor the
total mass, energy and momentum input to the ISM and IGM.

\section{Wind Physics}
\label{mechanisms}

There are three ways to accelerate a wind:

\noindent {\bf 1. Thermal gas pressure:} acts isotropically and has
  $v_{max} = v(sound) \sim$100~km~s$^{-1}$ for T$\sim$10$^7$K. Thermal
  pressure occurs naturally at inner edge of obscuring torus where
  temperatures of 10$^{6}$-10$^{7}$K are reached \cite{Krolik95}. But
  even slow winds have $v_r \sim 10~v(sound)$ and need to be driven by
  gas at $\sim$10$^9$K. Moreover, isotropic acceleration naturally
  and produces 100\% covering factors \cite{Balsara93}, yet half of
  all AGNs show no WAs or NALs. One clever way to heat gas is via
  cosmic rays. The decay of relativistic neutrons to protons a few
  parsecs from the nucleus can heat gas locally, without causing
  heating closer in \cite{Begelman91}.

\noindent {\bf 2. Radiation pressure:} acts radially and has $v_{max}
  = 2 v(critical point) \sim 2 v(Kepler, launch)$
  $\sim$10$^4$~km~s$^{-1}$. The radiation force has an 'Effective
  Eddington Limit' depending on the dominant mechanism: electron
  scattering (weak) \cite{King03}; UV line driving (which can be
  suppressed by strong X-rays overionizing gas) \cite{Murray95}; and
  dust absorption (which is only effective beyond the dust sublimation
  radius) \cite{Binette98}. UV line driving is the most discussed
  mechanism \cite{Murrayetal95, Proga00, Leighly04, Risaliti06}.

\noindent {\bf 3. Magneto-Centrifugal:} models accelerate plasma along
  field lines 'like beads on a wire', and has $v_{max} = c$. These
  winds remove angular momentum from the disk, which enhances
  accretion \cite{Blandford82, Kartje99}. A magneto-centrifugal base
  to a wind could provide the shielding gas for UV line driving
  further out \cite{Everett05}.

\smallskip
Radiation and magneto-centrifugal are the two most promising
mechanisms.  

\smallskip
Whichever mechanisms dominate in quasars, the wind must ultimately be
a function of the basic AGN parameters: the SMBH mass and the
accretion rates, both at the continuum emitting region (which drives
luminosity), and at the wind launching radius (which could limit the
wind mass supply).

The successful wind model will also have to explain the observed
regularities in quasar properties: the Baldwin effect
\cite{Baldwin77} - the luminosity dependence of the BEL equivalent
width - and 'Eigenvector 1' - a clustering of emission line and X-ray
properties that seems to be a function of \.{M}$_{acc}$
\cite{Marziani01}.

\section{Conclusions}

I have discussed why the details of the inner structure of quasars
makes for orders of magnitude differences in the strength of the
cosmological feedback from AGNs. The form of that feedback: energy,
momentum or mass, also depends on the details of the wind driving
mechanisms and obscuration.

Because slow, non-relativistic, winds are now thought to produce most
of the atomic emission and absorption features in AGN spectra a huge
range of possible tests of wind models has now opened up. With
adaptive optics poised to give diffraction limited near-infrared on
large telescopes, the dusty tori in AGN will be imaged down to parsec
scales, while smaller scale obscuration, possibly from a wind,
will be studied via X-ray and optical variability.

When we understand AGN winds we will not only have solved a
major part of AGN astrophysics, but have a strong basis for
extrapolating wind properties to all SMBHs at all redshifts. And then
we will be able to put AGN cosmological feedback on a firm basis.

\vspace{2mm}
{\em Acknowledgements:}
I am grateful to the other organizers of this meeting for the
invitation to speak, and to my session chair for letting me run over
by a factor 2 in time to respond to the questions that have improved
this paper. I depend on the expert assistance of my collaborators,
notably Yair Krongold, Nancy Brickhouse, Fabrizio Nicastro, Smita
Mathur, Guido Risaliti and Margarita Karovska. This work was funded in
part by NASA grant GO4-5126X ({\em Chandra}).



\end{document}